# A direct time measurements technique for the two-dimensional precision coordinate detectors based on thin-walled drift tubes

A.M. Makankin[1], V.V. Myalkovskiy[1], V.D. Peshekhonov[1], S. Ritt[2], S.E. Vasilyev[1]

[1] Joint Institute for Nuclear Research (JINR), Dubna, Russia
[2] Paul Scherrer Institute (PSI), Villigen, Switzerland

**Abstract**

This article presents the results of a study of the longitudinal spatial resolution of 2 m long straw tubes by means of the direct timing method (DTM). The feasibility of achieving a coordinate resolution (r.m.s.) better than 2 cm over full length of the straw is demonstrated. The spatial resolution insignificantly changes when measured by detecting gammas from a Fe-55 gamma ray source or minimum ionizing particles from a Ru-106 source. The use of the same type of FEE for data taking both for measuring the drift time of ionization electrons and propagation of a signal along the anode wire allows one to construct a two-dimensional detector for precision coordinate measurements.

## 1. Introduction

The measurement of the momentum of charged particles moving in a magnetic field requires precise measurements of the spatial coordinates in the direction of deflection, whereas the second coordinate which determines the angle of the incident track can be measured with smaller precision. A high precision coordinate is determined in gaseous drift chambers in the direction orthogonal to the anode wires by means of measurements of the drift time of the ionization electrons produced by charged particles passing through the chamber. A few different methods can be employed for the measurements of the coordinate along the anode wire. They include a cathode strip readout [1, 2], and two coordinate drift chambers with a pad read out which are under construction [3, 4]. Both these techniques can be applied for the coordinate chambers based on thin walled drift tubes (straw tubes) but in this case the cathodes should have a resistance higher than 100 K$\Omega$/□. Moreover, the charge collection time should be increased as well [1-3]. The longitudinal coordinate in the straw can be also determined by employing the charge-division method when the signals are read out from two ends of the resistive anode wire. The highest longitudinal resolution achieved by this method has been obtained with the help of auxiliary measurements involving a time-charge asymmetry [5, 6]. The anode resistivity was 400 $\Omega$/m. The best spatial resolution $\sigma$ proved to be 0.95 cm in the straw center and about 2.5 cm near the ends of the 1.52 m straw if 5.9 keV gammas were registered. However, the resolution deteriorated down to 2.5 and ~6.0 cm, respectively, if minimum ionizing particles were registered.

Earlier, a possibility of registering a longitudinal coordinate in a drift-tube detector by employing technique of measurements of a time difference between signals arriving to amplifiers connected to the ends of the anode wire has been demonstrated [7]. The technique has been called a direct timing method (DTM). In this case, the anode wire is considered as a transmission line.

In the present article we consider possibilities of using the DTM for registering signals from the coordinate detector prototype constructed of 2 m long straw tubes. The particular features of the prototype include a possibility of operation at the gas-mixture pressure up to 4 bar and the special design of the gas distribution manifolds which allows one to assemble the readout front-end electronics close to the ends of anode wires.

## 2. The Setup

The detector prototype includes a layer of glued straw tubes with an inner and outer diameter of 9.53 and 9.65 mm, respectively. The 2 m long straw tubes are produced by winding two Kapton strips. The inner strip is a carbon loaded Kapton conductive film of XC-160 type, whereas the outer strip is Kapton aluminized film of HN50 type [8]. The anode is made of 30 µm gold-plated tungsten wire which has a resistance of 70 $\Omega$/m and installed under the tension of 70 g with one polycarbonate spacer at the middle of the straw tube. The transmission line impedance is 360$\Omega$. The straw tubes were flashed with the gas mixture Ar/$CO_2$ (80/20) at the pressure of 1 or 3 bar. The layout of the experimental setup is shown in Fig. 1. A straw under study has been irradiated either with 5.9 keV gammas from a Fe-55 gamma ray source or with beta rays from Ru-106 source through the slit collimators. In the latter case, the electrons passing through the straw have been detected by a scintillation counter with two PM. Low-energy electrons have been stopped in an absorber located between the straw and the counter.

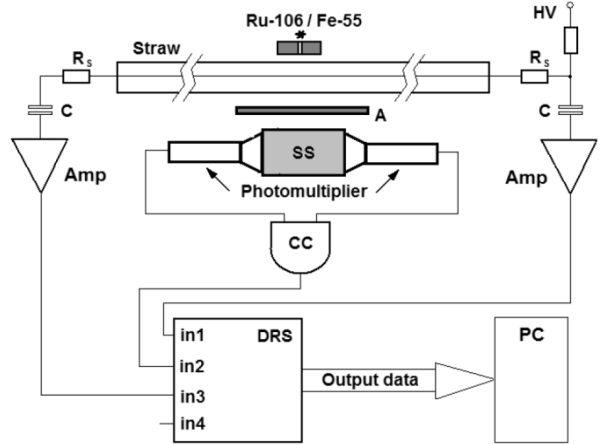

Fig. 1. A schematic view of the setup: A – absorber, SC – scintillation counter, Amp – amplifier, CC – coincidence circuit, DRS4 – Domino Ring Sampler (DRS).

Amplifiers based on a MSD-2 microcircuit with a gain of 35 mV/μA and a rise time of about 4 ns, an input impedance of 120Ω [9], are installed close to the ends of the anode wire connected with minimum parasitic capacitance and inductance. Such amplifiers are used in detectors when a radial coordinate is determined by measuring the drift time of the ionization electrons [10].

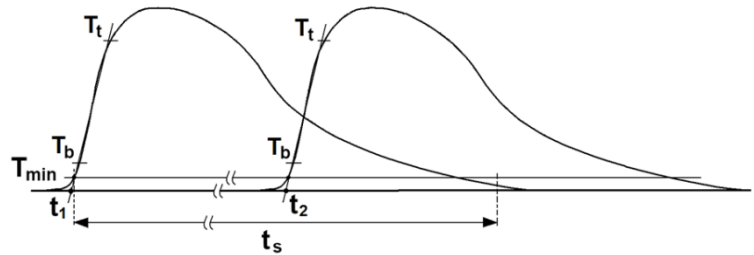

Fig. 2. A schematic representation of the developed algorithm.

The pulses from the outputs of the amplifiers are fed into two channels (1 and 3) of the DRS4 Evaluaiton Board. This board is based on the DRS4 Switched Capacitor Array chip (SCA), which can sample an input signal with a sampling speed of up to 5 GS/s and store an analog waveform in a time window of the size determined by the capacitor array of the DRS4 chip (1024 words – 200 ns) [11, 12]. The maximal amplitude of the input pulses is ±0.5 V. The data arriving at DRS4 is processed according to the following algorithm (Fig. 2):
- selection of the correlated pulses by introducing a threshold for the incoming anode pulses, and by requesting a signal from the scintillation counter arriving at the second DRS4 channel in the case of measurements with the Ru -106 source;
- determination of the pulse amplitudes in the software time window defined as $t_s$ and normalization of the delayed pulse height on the preceding pulse amplitude;

- linear approximation of the leading edge in the time interval between $T_t$ and $T_b$ levels determined by the software as a percentage of the highest pulse amplitude;
- determination of the arrival time for each pulse ($t_1$ and $t_2$) as the intersection point of the approximate line and the time axis;
- plotting a histogram of the time difference of the delay between pulses $\Delta t=(t_1-t_2)$ and evaluation of the mean and standard deviation to determine the resolution of the method for longitudinal avalanche coordinates along the anode.

If the point of the avalanche origin is displaced along the anode wire by δ*y* from its middle the two signals arriving at amplifiers pass the distance L/2 ± δ*y*, where L is the anode length. Therefore, the absolute difference δτ is determined as δτ = 2δ*y*/v where v is an electromagnetic wave propagation velocity along the anode wire whereas the sign determines the direction of the coordinate displacement with respect to the center of the wire.

The DRS4 linearity proved to be satisfactory (Fig. 3). It has been verified by feeding a pair of pulses with a calibrated delay varied in the time interval from 2 to 70 ns. The distribution of the time interval between the pulses has demonstrated that the apparatus resolution σ is within the range of 60–85 ps. This uncertainty arises from the design principle of the DRS4 chip. The sampling clock is generated by a pulse propagating through an inverter chain. Imperfections in the chip production cause a variation of the individual delays of the inverters, causing a timing non-linearity. Since these imperfections are stable over time, they can be measured and removed by calibration. The currently used calibration method improves the timing resolution from ~600 ps to the above mentions value. It has recently been shown [9] that new algorithms are able to improve this further by a factor of about three. The new calibration will soon be implemented and an improved measurement will be pursued.

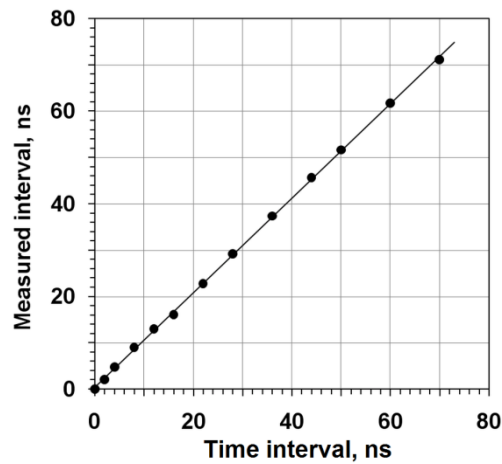

Fig. 3. The measured time interval as a function of the calibrated delay interval varied in the range of 2–70 ns (DRS4 linearity).

**Parameters of the detected signals**

A Fe-55 gamma ray source has been used both in the studies of the DTM technique and in the preliminary measurements. The pulses from the amplifiers connected to the ends of the anode wire proved to be identical (Fig. 4a) if the source position was at the center of the straw. Then, the scan along the straw allowed for the measurements of the attenuation in the pulse amplitude and the pulse propagation velocity along the anode wire. Typical pulses registered by an amplifier are shown in Fig. 4b for the case when the gamma ray source is located at the end of the straw tube. The attenuation of the pulses as a function of the distance between the source and the straw end is shown in Fig. 4c. Apparently, the pulse attenuation at the length of about 2 m is 1.17 and the pulse propagation velocity was found to be is 3.49 ns/m.

We have considered the pulse waveforms from sources Fe-55 and Ru-106 for different termination of the straw with the amplifiers. The different measurements included an option without series resistors $R_s$ (Fig.1) at the input of the amplifier with the impedance of 120Ω as well as the options with resistors of 180 and 240Ω connected in series to the input. As a result, the leading edge of the signals measured at the pulse height 0.1—0.9 varies from 6 to 11 and to 15 ns, respectively. Imperfect termination in the first two cases resulted in a considerable change in the pulse waveforms caused by superposition of the reflected signal thus hampering optimization of $T_t$ and $T_b$ levels along the straw. Some signal shapes from the gamma ray source for different termination options are displayed in Fig. 5.

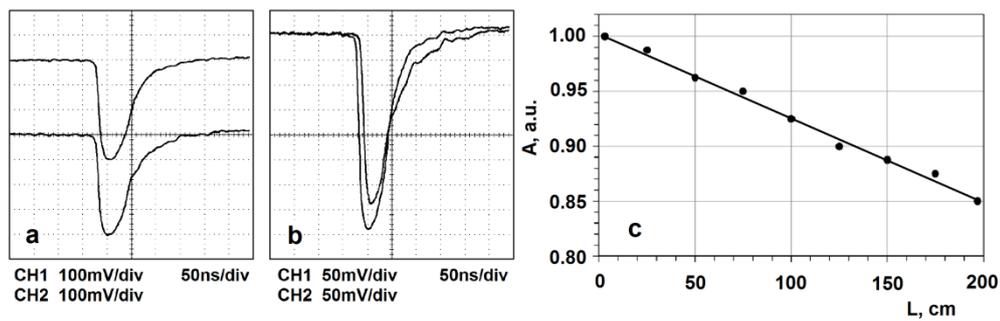

Fig. 4. (a) – Signals registered at the ends of the anode when a collimated Fe-55 source is located at the middle of the straw; (b) – Typical signals registered by an amplifier when a collimated Fe-55 source is located near the end of the straw; (c) – Signal attenuation as a function of the distance from the straw end.

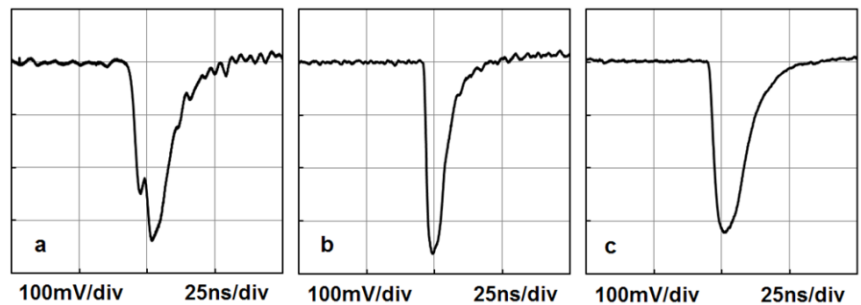

Fig. 5. Typical output signals obtained with a Fe-55 source: (a) – without a series resistor Rs;
(b) – with Rs = 180Ω; (c) – with Rs = 240Ω.

When two signals are registered at the ends of the straw tube, the distance traversed by the wave along the anode wire is different for the two ends which results in different changes of the pulse waveforms and different slope of the leading edge of these pulses. The difference between the changes of the leading edge of the fast and delayed pulses ($\tau_1 - \tau_2$) is shown in Fig.

6 as a function of the distance of the Fe-55 source from the middle of the straw. The changes are displayed in Fig. 6a and 6b for the termination of 120 and 360Ω, respectively, in the range from 0.15 to 0.75 of the pulse height of the preceding pulse. The difference at the level of 0.15 was minimal and assumed to be zero. One can see that the changes in the leading edge is insignificant if the difference in the signal propagation distance along the anode wire is about 20–30 cm, which corresponds to displacement of the source by 10–15 cm away from the straw center. The difference in time increases substantially both with increasing of the displacement from the center and of the level at the leading edges.

Large dynamical range of the signals produced by minimum ionizing particles as well as the clustering of the ionizing losses lead to greater distortions of the pulse waveform compared to those produced by the Fe-55 source. These distortions can result in a nonlinearity in the determination of the pulse arriving time if approximate curves are used. The accuracy of the method is therefore dependent on the data analysis procedure. A few pulses registered for the straw tube termination of 360Ω are displayed in Fig. 7.

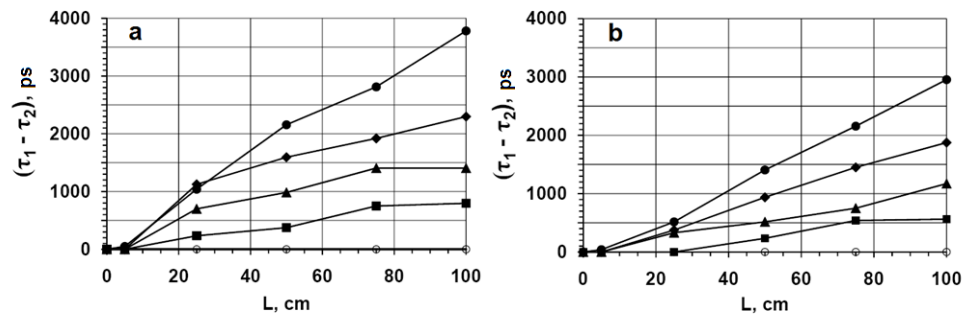

Fig. 6. A time difference between leading edges of the pair of pulses as a function of the distance of a Fe-55 source from the middle of the straw tube: (a) – termination of the straw is 120Ω; (b) – termination is 360Ω. The symbols correspond to different signal levels at the leading edge as a fraction of the peak value: 0.15 – open circles, 0.3 – filled squares, 0.45 – filled diamonds, and 0.75 – filled circles. Gas gain (G) is ~$4 \times 10^4$.

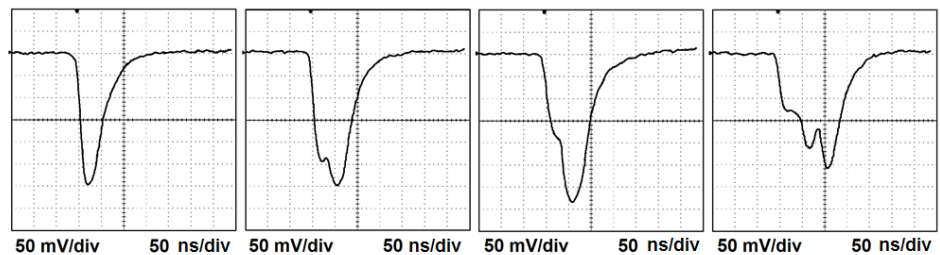

Fig. 7. Typical pulses obtained by detection of high energy electrons from a Ru-106 source. The termination is 360Ω, the anode voltage is 1.85 kV (G is ~$4.5 \times 10^4$).

## 4. Longitudinal spatial resolution

### 4.1. Gamma-ray detection resolution

In the method of the charge division the longitudinal spatial resolution is determined mainly by the signal-to-nose ratio which deteriorates substantially due to passage of the signal through the high-resistive anode wire. The direct timing method allows one to use low resistive wires usually employed in the gaseous drift detectors as anode. Both the low dynamic range and lower pulse shape distortion in registering gammas compared to minimum ionizing particles indicate to possibility of reaching better longitudinal resolution in the detection of gammas from a Fe-55 gamma ray source. The changing of the longitudinal resolution along the straw was studied by moving the Fe-55 source from the center of the straw to its end. The algorithm employed for the data processing included the following parameters. The threshold was 10 mV. The time window $t_s$ used in determining of the pulse heights started at the threshold level at the pulse leading edge and had a length of 400 bins of the clock generator. The levels of $T_t$ and $T_b$ have value as 0.1 and 0.7 of the fast pulse amplitude, respectively. The time shift $\Delta t$ between two pulses for each pair is used for the determination of the longitudinal resolution. We investigated a possibility to obtain the best resolution by optimization of the levels $T_t$ and $T_b$ for both pulses and by variation of the $\Delta t$ in the range of $\pm 3$ bins.

Curves in Fig. 8 show the longitudinal resolution obtained with the optimization of the algorithm parameters (●) or without optimization (▲) at different positions of the source along the straw tube and at different gas pressures (8-a and 8-b). In both cases one can reach the longitudinal resolution in the interval from ~70 to 125 or 210 ps and from ~50 to 100 or 210 ps, for the gas pressure 1 and 3 bar, respectively. Note, that 1 cm distance corresponds to the difference in the time delay $\Delta t = 69.8$ ps.

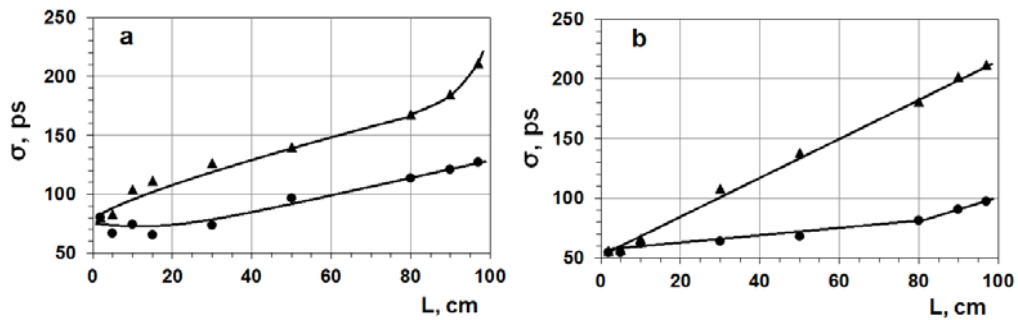

Fig. 8. Longitudinal resolution along the straw as a function of the distance of a Fe-55 source from the straw center: filled circles – the highest resolution obtained with optimization of the parameters of the data processing, filled triangles – the resolution obtained with fixed parameters $\Delta \tau$, $T_b$ и $T_t$. Zero value of the abscissa corresponds to the center of a straw. The termination is 360Ω. (a) – The gas mixture pressure is 1 bar and the anode voltage is 1.85 kV, gas gain is $4 \times 10^4$. (b) – The gas mixture pressure is 3 bar and the anode voltage is 3.05 kV.

## 4.2. Spatial resolution in the case of electron registration

Some pulses obtained in the case of registration of electrons from the Ru-106 source shown in Fig. 7 indicate to potential distortions of the measurements of time shift $\Delta t$ between two pulses. In view of this problem, one should develop a special processing algorithm even in the case of an optimal termination of the straw. The highest longitudinal resolution obtained for the straw at a gas pressure of 1 bar with the termination of 360Ω is shown in Fig. 9 (curve 1). The resolution varies from ~120 ps in the center of the straw to ~160 ps over ~80% of the full straw tube length, and deteriorates to ~180 ps near the end. Measurement of longitudinal resolution for fixed parameters of the processing (curve 2) showed a deterioration of the resolution from 150 ps at the center of the straw up to ~280 ps at its end.

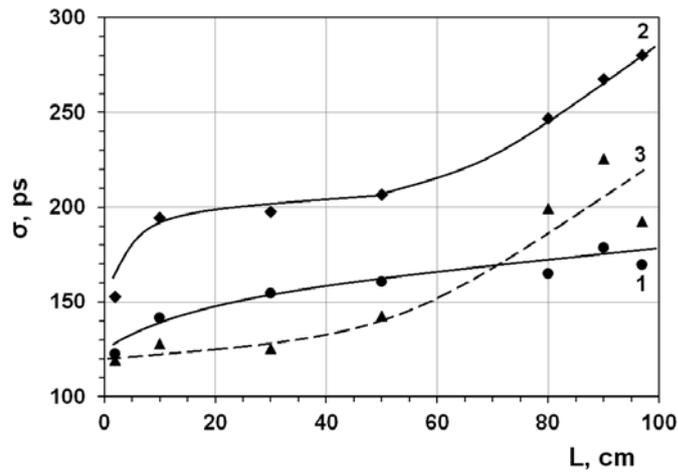

Fig. 9. The highest longitudinal resolution along the straw as a function of the distance of the Ru-106 source from the straw center. Curve 1 corresponds to the highest resolution. The resolution obtained for the fixed parameters of the data processing is shown with curve 2. The resolution for the fixed parameters with taking in account the adjustment values of $\Delta N_b$ (Fig. 10) is shown by curve 3. The gas mixture pressure is 1 bar and the anode voltage is 1.95 kV. The termination of the straw is 360Ω, and the gas gain is about $8 \times 10^4$.

To provide the spatial resolution close to the highest of its values along the straw should be introduced correction of the measured interval $\Delta t$ and other improving. The measured value of $\Delta t$ is expressed as $\Delta t = 0.2 \cdot (N_b - 1) + \delta t$, where $N_b$ is the number of full 200 ps bins and $\delta t$ is a part of the last bin. For example, the longitudinal resolution at a fixed position of the source for the changing of the measured value $\Delta t$ at several units of the bins is shown in Figure 10. In this case the changing of the $\Delta t$ from its measured value on $\Delta N_b$ bins may improve the resolution from ~280 to 190 ps. The coordinate dependence of the longitudinal resolution shown by curve 3 in Fig. 9 corresponds to the data processing with the fixed parameters and with the using of the optimal $\Delta N_b$ values for these locations of the source. One can see the improvement of the resolution, which for the two meter straw is in the range from 120 to ~230 ps.

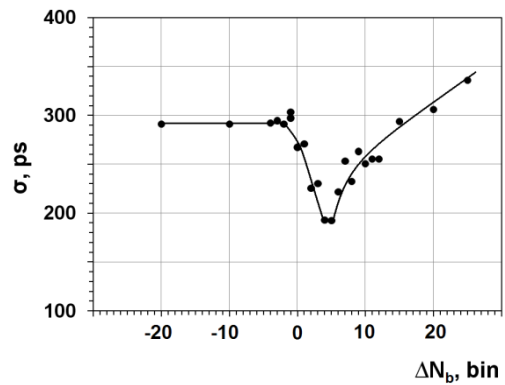

Fig.10. The longitudinal resolution as a function of the variation of the $\Delta N_b$ near the measured value of $N_b$. The distance of a Ru-106 source from the straw center is 97 cm. Position of 0 (X-axis) corresponds to the primary measured value $N_b$.

The correlation between the calculated values of the timing delayed intervals of $\Delta t_c$ for the different position of the source along the straw and the obtained values of these intervals of $\Delta t_m$ is displayed in Fig. 11. The fit yields $\Delta t_m = 1.3 \Delta t_c$, that should be considered when processing real data.

Different attenuation of the fast and delayed pulse gives different changes of the pulse waveforms and different slope of these leading edges, which increases the measured value of $\Delta t$ in dependent on the distance of the avalanche from the center of the straw. The changing of the time scale by the factor 1.3 demands an adjustment of the difference between the measurement time delay $\Delta t_m$ and calculated time. Evidently, that 1 cm distance corresponds to the measured difference in the time delay $\Delta t_m = 90.7$ ps. It is this value that should be used for a conversion of the longitudinal resolution to the units of cm. In that way, the longitudinal resolution shown by curve 3 in Fig. 9 corresponds to the full range of the coordinate resolution from ~1.3 to less than 2 cm for 80% of the straw length, and better than 2.6 cm for the remaining 20% of the length.

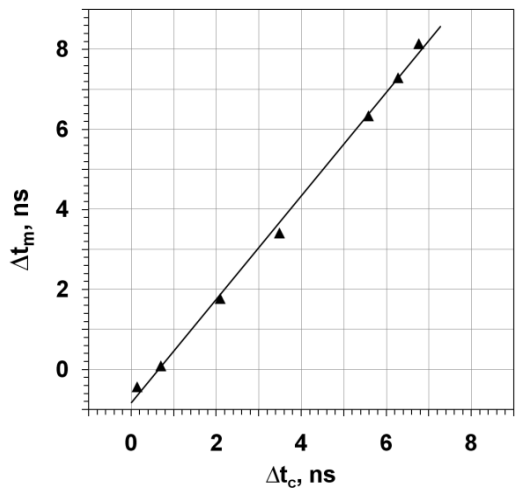

Fig. 11. The correlation between the measured ($\Delta t_m$) and the calculated ($\Delta t_c$) time differences along the straw.

## 5. Conclusions

The measurement conducted with the detector prototype have demonstrated feasibility of using the direct timing method for the determination of the longitudinal coordinates in the 2 m long thin-walled drift tubes to the accuracy less than 2 cm ($\sigma$). Fast current amplifiers can be used both for radial coordinate measurements which rely on the measurements of the electron drift time in the direction orthogonal to the anode and for the coordinate measurements in the longitudinal direction by the DTM technique. Thus, they can serve as amplifiers for two-dimensional fast coordinate detectors.

We have demonstrated that the coordinate information can be obtained for the full straw tube length by fixing parameters, which gives the possibility of using the DTM technique in multichannel detectors. In this case, the longitudinal resolution of 2 m long tube straw is now three times better in comparison with the charge division method. The DTM provides about 2 cm of the longitudinal resolution for the 1.5 meters of the tubes instead of 6 cm provided by the charge-division method. At the same time one finds that the developed algorithm should be modified in order to achieve the highest possible resolution along the full straw tube length. A more uniform longitudinal resolution at the full straw tube length will be achieved by the improved timing calibration of the DRS4 chip in the near future. These and some other modifications of the algorithm should increase the uniformity of the resolution along the straw tube and make the longitudinal resolution closely to 2 cm or better for longer tubes.